\documentclass[aps,prd,twocolumn,floats,amsmath,preprintnumbers,nofootinbib,superscriptaddress,amssymb,10pt]{revtex4-2}

\usepackage{graphicx}
\usepackage{dcolumn}
\usepackage{bm}
\usepackage[hidelinks]{hyperref}
\usepackage{cleveref}
\usepackage{xcolor}
\usepackage{physics}
\usepackage{multirow}
\usepackage{amsfonts}
\usepackage{subcaption}
\usepackage[normalem]{ulem}
\usepackage{braket}
\usepackage{orcidlink}

\usepackage[font=small, labelfont=bf, textfont={small,it}]{caption}
\graphicspath{{figs/}}

\renewcommand\thesection{\arabic{section}}

\newcommand{\beq}{\begin{eqnarray}}
\newcommand{\eeq}{\end{eqnarray}}
\newcommand{\beqnn}{\begin{eqnarray*}}
\newcommand{\eeqnn}{\end{eqnarray*}}

\newcommand{\Var}{\ensuremath{\mathrm{Var}}}

\newcommand{\SU}{\mathrm{SU}}

\newcommand{\cool}{\mathrm{cool}}

\newcommand{\nstep}{n_{\mathrm{step}}}

\newcommand{\ndof}{n_{\mathrm{dof}}}
\newcommand{\nbetween}{n_{\mathrm{between}}}

\newcommand{\tauint}{\tau_{\mathrm{int}}}
\newcommand{\DKL}{\tilde{D}_{\mathrm{KL}}}
\newcommand{\ESS}{\mathrm{ESS}}
\newcommand{\hESS}{\hat{\mathrm{ESS}}}

\newcommand{\Pf}{\mathcal{P}_{\mathrm{f}}}
\newcommand{\Pre}{\mathcal{P}_{\mathrm{r}}}
\newcommand{\Wd}{W_{\rho,\mathrm{d}}}

\newcommand{\neavg}[1]{\langle #1 \rangle_\mathrm{f}}

\begin{document}

\setlength\abovedisplayskip{10pt}
\setlength\belowdisplayskip{10pt}

\setlength{\parskip}{4pt}
\setlength{\parindent}{0pt}

\preprint{}

\title{Ergodic topology sampling toward the continuum with Stochastic Normalizing Flows}

\author{Claudio Bonanno~\orcidlink{0000-0002-1421-6510}}
\email{claudio.bonanno@unibe.ch}
\affiliation{Albert Einstein Center for Fundamental Physics, Institute for Theoretical Physics, University of Bern, Sidlerstra{\ss}e 5, CH-3012 Bern, Switzerland}

\author{Andrea Bulgarelli~\orcidlink{0009-0002-2917-6125}}
\email{abulgare@uni-bonn.de}
\affiliation{Transdisciplinary Research Area ``Building Blocks of Matter and Fundamental Interactions'' (TRA Matter) and Helmholtz Institute for Radiation and Nuclear Physics (HISKP), University of Bonn, Nussallee 14-16, 53115 Bonn, Germany}

\author{Elia Cellini~\orcidlink{0000-0002-5664-9752}}
\email{elia.cellini@ed.ac.uk}
\affiliation{Higgs Centre for Theoretical Physics, School of Physics and Astronomy,\\
The University of Edinburgh, Edinburgh EH9 3FD, United Kingdom}

\author{Alessandro Nada~\orcidlink{0000-0002-1766-5186}}
\email{alessandro.nada@unito.it}
\affiliation{Dipartimento di Fisica,  Universit\`a degli Studi di Torino and INFN, Sezione di Torino, Via Pietro Giuria 1, I-10125 Turin, Italy}

\author{Dario Panfalone~\orcidlink{0009-0007-6651-7490}}
\email{dario.panfalone@unito.it}
\affiliation{Dipartimento di Fisica,  Universit\`a degli Studi di Torino and INFN, Sezione di Torino, Via Pietro Giuria 1, I-10125 Turin, Italy}

\author{Davide Vadacchino~\orcidlink{0000-0002-5783-5602}}
\email{davide.vadacchino@plymouth.ac.uk}
\affiliation{Centre for Mathematical Sciences, University of Plymouth, Plymouth, PL4 8AA, United Kingdom}

\author{Lorenzo Verzichelli~\orcidlink{0009-0008-0825-4845}}
\email{lorenzo.verzichelli@unito.it}
\affiliation{Dipartimento di Fisica,  Universit\`a degli Studi di Torino and INFN, Sezione di Torino, Via Pietro Giuria 1, I-10125 Turin, Italy}
\affiliation{Present address: Fakult\"at f\"ur Physik, Universit\"at Bielefeld, Universit\"atsstra\ss e 25, D-33615 Bielefeld, Germany}
\date{\today}

\begin{abstract}
\noindent In this work we provide numerical evidence that topological freezing in $\mathrm{SU}(3)$ Yang--Mills theory can be mitigated with a flow-based approach, as tested at lattice spacings as fine as 0.03~fm. We do so by analyzing the scaling of the costs of a computational strategy based on the combination of open boundary conditions and Stochastic Normalizing Flows. We show that the integrated autocorrelation time of the squared topological charge and the statistical uncertainty on the topological susceptibility remain under control throughout the explored parameter range with currently available computational resources. This marks an important step toward the use of even more efficient flow-based architectures for ergodic sampling of topology in state-of-the-art lattice gauge theory calculations.
\end{abstract}

\maketitle

\section{Introduction}\label{sec:intro}

Theoretical and phenomenological implications of gauge topology are among the most interesting non-perturbative features of Yang--Mills theories and Quantum Chromodynamics (QCD). The existence of a non-trivial topological charge $Q=\frac{1}{64\pi^2}\varepsilon_{\mu\nu\rho\sigma}\int \dd^4 x \, G^a_{\mu\nu}(x)G^a_{\rho \sigma}(x)$, which labels different topological sectors of the gluon field space, is responsible, e.g., for the anomalous breaking of the axial symmetry~\cite{tHooft:1976rip}, and for the non-trivial dependence on the $\theta$-parameter~\cite{Jackiw:1976pf}. These properties are intimately connected with several phenomenological aspects of hadron physics and beyond, such as the $\mathrm{U}(1)_{\mathrm{A}}$ puzzle and the large $\eta^\prime$ mass~\cite{Witten:1979vv,Veneziano:1979ec,Witten:1980sp, DiVecchia:1980yfw}, the strong CP problem and the axion~\cite{Peccei:1977hh,Peccei:1977ur,Wilczek:1977pj,Weinberg:1977ma}, or the chiral magnetic effect in heavy-ion collisions~\cite{Kharzeev:2007jp,Fukushima:2008xe,Fukushima:2010vw,Kharzeev:2013ffa}.

Due to their intrinsic non-perturbative nature, numerical Markov chain Monte Carlo (MCMC) simulations on the lattice are a natural tool to investigate topological features of gauge theories from first principles. However, this is notoriously challenging due to the infamous \emph{topological freezing} problem~\cite{Alles:1996vn,DelDebbio:2002xa,DelDebbio:2004xh,Schaefer:2010hu}. At coarse lattice spacing $a$, topological sectors are not well defined, and the Markov Chain used to sample the lattice path integral distribution is able to easily jump from one sector to the other, ensuring ergodicity. When the lattice spacing is shrunk to zero to approach the continuum limit $a\to 0$, a proper definition of topological sectors is eventually restored~\cite{Luscher:1981zq}. As energy barriers between would-be topological sectors grow (and eventually diverge), it becomes increasingly difficult for the Markov Chain to jump from one sector to another when standard local algorithms are employed. At very fine lattice spacing, few to no jumps are observed, and eventually the system gets trapped in a fixed topological sector, causing a freezing of the Monte Carlo evolution of $Q$. This implies a severe loss of ergodicity, threatening one of the main assumptions underlying the calculation of expectation values in lattice field theories using MCMC methods.

The severity of the freezing is quantitatively tracked by the growth of the integrated autocorrelation time $\tau_{\rm int}$ of some power of $Q$ (a typical choice is $Q^2$) as a function of $1/a$. This growth can either be described by a power law with a much larger dynamical critical exponent compared to the one of local observables (e.g., small Wilson loops), or by an exponential growth, see, e.g.,~\cite{DelDebbio:2004xh,Schaefer:2010hu,Bruno:2014ova,Bonati:2015sqt,Eichhorn:2023uge}. 

Topological freezing is not only a serious computational problem that prevents the correct sampling of the path integral distribution of the topological charge (and thus the calculation of important topological observables such as the topological susceptibility), but can also potentially bias the calculation of any expectation value, in the form of finite-size effects suppressed as a power-law (as opposed to exponentially)~\cite{Brower:2003yx, Aoki:2007ka}. 
The relevance of this systematic effect depends in practice on how large the coupling of topological modes to the observable of interest is: relevant examples in this respect are the energy density after the gradient flow~\cite{Fritzsch:2013yxa,RamosMartinez:2023tvx}, or reference flow times such as $t_0$ used for scale setting~\cite{Bergner:2014ska,Bonanno:2025kfd}.

For these reasons, in the last few years algorithmic development in lattice field theory to tackle topological freezing
has received a lot of attention. Several directions are being pursued by the community, see, e.g.,~\cite{Bietenholz:2015rsa,Laio:2015era,Luscher:2017cjh,Bonati:2017woi,Giusti:2018cmp,Florio:2019nte,Funcke:2019zna,Albandea:2021lvl,Cossu:2021bgn,Borsanyi:2021gqg,Fritzsch:2021klm,Eichhorn:2023uge,Howarth:2023bwk,Albandea:2024fui,Abe:2024fpt,Holland:2024muu,Holland:2025fsa,Katz:2025ycr,Eichhorn:2026eok,Blossier:2026itg}, or Refs.~\cite{Finkenrath:2023sjg,Boyle:2024nlh,Finkenrath:2024ptc} for recent reviews.  
A number of them build on the well-known improvement in topology autocorrelations yielded by the adoption of open boundaries~\cite{Luscher:2011kk,Luscher:2012av,McGlynn:2014bxa}, an idea that has been applied quite extensively in recent years in the context of the tempering in boundary conditions. Combining open and periodic boundaries has the benefit of lifting the improved scaling of autocorrelations from the open to the periodic system, where calculations can be performed free of systematic boundary effects. This strategy, implemented by means of parallel tempering schemes, has been shown to yield an improved scaling of simulation costs with respect to both purely periodic and purely open boundary simulations~\cite{Hasenbusch:2017unr,Bonanno:2020hht,Bonanno:2024zyn,Bonanno:2025eeb}.

More recently, Non-Equilibrium Monte Carlo (NEMC) simulations have also exploited tempering schemes in the boundary conditions for the mitigation of topological freezing~\cite{Bonanno:2024udh,Bonanno:2025pdp}, leveraging Jarzynski equality~\cite{Jarzynski:1996oqb,Jarzynski:1997ef,Jarzynski:1998ef}. In the last decade the latter approach has been extensively applied to the computation of ratios of partition functions for a large set of applications in lattice field theory~\cite{Caselle:2016wsw,Caselle:2018kap,Francesconi:2020fgi,Bulgarelli:2023ofi, Bulgarelli:2024onj,Bulgarelli:2025riv}. In the context of lattice sampling methods, instead, the adoption of NEMC is particularly interesting because it belongs to a broader class of algorithmic strategies that has been extensively investigated in the last years, i.e., \emph{flow-based} approaches.

Initially proposed in the context of trivializing maps~\cite{Luscher:2009eq,Engel:2011re}, flow-based approaches have more recently re-emerged as promising alternatives to standard MCMC sampling with the development of Normalizing Flows (NF) for lattice field theory~\cite{Albergo:2019eim,Nicoli:2020njz}. NFs are a class of generative models~\cite{rezende2015,papamakarios2021} powered by deep neural networks capable of parameterizing the highly complex transformations needed to connect one tractable probability distribution (the prior) with the desired target one. They are natural tools to circumvent critical slowing down in a general fashion by avoiding direct sampling of the target distribution altogether; at the same time, asymptotic exactness is ensured by appropriate reweighting or accept-reject procedures. After initial applications to scalar field theories~\cite{DelDebbio:2021qwf,Gerdes:2022eve,Albandea:2023wgd}, a wide community effort focused on finding efficient NF architectures for lattice gauge theories~\cite{Kanwar:2020xzo,Boyda:2020hsi,Bacchio:2022,Abbott:2023thq,Gerdes:2024rjk} with the inclusion of fermions as well~\cite{Finkenrath:2022ogg,Albergo:2022qfi}.\footnote{For applications to other physically interesting lattice field theories see Refs.~\cite{Caselle:2023mvh,Kreit:2025cos,Schuh:2025gks}.} We point to Ref.~\cite{Cranmer:2023xbe} for a review and to Refs.~\cite{Albergo:2021vyo,Cheng:2025khv} for an introduction to the matter.

The minimization procedure (the ``training'') needed to determine the highly complex transformations at the core of NF architectures has proved to be challenging for higher-dimensional systems, in particular in terms of the scaling of training costs \cite{DelDebbio:2021qwf,Abbott:2022zsh}. Due to these difficulties, NFs are not (yet) a viable strategy to challenge traditional sampling techniques in state-of-the-art calculations in four-dimensional gauge theories.\footnote{For recent breakthroughs powered by NFs in variance reduction techniques see, e.g., Refs.~\cite{Abbott:2024kfc,Abbott:2026ylv,Abbott:2026vui}.}

In the context of flow-based samplers, NEMC provides a purely stochastic approach, to be contrasted to the deterministic nature of NF architectures: it is however extremely natural to combine them by alternating (non-equilibrium) MCMC updates with appropriate transformations of the fields. This is the structure of the recently introduced Stochastic Normalizing Flows (SNFs)~\cite{wu2020stochastic,Caselle:2022acb}, the flow-based architecture at the backbone of this work (see also Refs.~\cite{Vaikuntanathan_2011,Nilmeier_2011} for related earlier ideas). On the one hand, SNFs retain the same favorable scaling properties of NEMC sampling with respect to the relevant degrees of freedom, see Refs.~\cite{Bonanno:2024udh,Bulgarelli:2024brv,Bonanno:2025pdp}. On the other hand, they systematically improve NEMC by reaching the same efficiency with a fraction of the sampling costs and with small or negligible training costs. For a selection of applications of SNFs, see also Refs.~\cite{Caselle:2024ent,Bulgarelli:2024yrz,Kreit:2026eng}; for closely related non-equilibrium-based architectures see Refs.~\cite{Albergo:2024trn,Chen:2026bqw}. Finally, we would like to note that there is another transport strategy which has seen recent attention in the lattice community, which is also based on a stochastic process, namely diffusion models. These architectures are able to transform configurations between different distributions learning appropriate Langevin processes\footnote{An important distinction is that NEMC and SNFs define an explicit path between the distributions, whereas diffusion model do that only implicitly.}: recent works highlighted the connection with stochastic quantization~\cite{Wang:2023exq} and implemented these architectures in lattice gauge theories~\cite{Zhu:2025pmw,Vega:2025hgz,Aarts:2026zzr,Alharazin:2026lcb,Komijani:2026lan,Tomiya:2026tbc,Rothkopf:2026apb}. 

The goal of this paper is to present the first scaling study of SNFs in realistic conditions down to very fine lattice spacings, and to show that both integrated autocorrelation times for $Q^2$ and statistical uncertainties on $\braket{Q^2}$ are under control in the explored parameter range. To our knowledge, this is the first such study in the context of topological freezing mitigation for a flow-based sampling algorithm down on available computational resources for a four-dimensional non-Abelian gauge theory. As such, it represents an important step forward in the ongoing program of leveraging flow transformations and annealing procedures for the construction of efficient and ergodic sampling techniques close to criticality.

\section{Theoretical and numerical setup}
\label{sec:setup}

The pure $\SU(3)$ Yang--Mills theory is discretized on an hypercubic $L^4$ space-time lattice with lattice spacing $a$ using the Wilson plaquette action. Periodic boundaries are assumed everywhere except for the temporal gauge links stemming from an $L_{\mathrm{d}}^3$ cubic defect on the temporal boundary: in this case their gauge coupling is multiplied by a factor $0\le \lambda_n \le 1$, i.e.,  $\beta \to \lambda_n\beta$, which is varied from 0 (open boundary conditions, OBC) to 1 (periodic boundary conditions, PBC) during the out-of-equilibrium trajectory. The idea is to evolve the system out of equilibrium from open to periodic boundaries by changing $\lambda_n$ in $\nstep$ steps. Here, $n$ labels the intermediate distributions connecting the prior and the target ones (with open/periodic boundaries, respectively). At the $n^{\rm th}$ out-of-equilibrium step, the action reads:
\begin{equation}
S_{\lambda_n}(U^{(n)}) = - \frac{\beta}{N} \sum_{x, \mu < \nu}  K_{\mu\nu}^{(n)}(x)\Re\Tr\left[P_{\mu\nu}^{(n)}(x)\right],
\end{equation}
with $P_{\mu\nu}^{(n)}(x)=U^{(n)}_\mu(x)U^{(n)}_\nu(x+\hat{\mu}) {U^{(n)}_\mu}^\dagger(x+\hat{\nu}) {U^{(n)}_\nu}^\dagger(x)$ the plaquette with its corner in the site $x$ in the $(\mu,\nu)$ plane built in terms of the gauge links $U^{(n)}$, $\beta=2N/g^2$ the inverse bare coupling with $N=3$ the number of colors, and
\begin{equation}
K_{\mu\nu}^{(n)}(x) = K^{(n)}_\mu(x) K^{(n)}_\nu(x+\hat{\mu}) K^{(n)}_\mu(x+\hat{\nu}) K^{(n)}_\nu(x),
\end{equation}
where $K^{(n)}_\mu(x)=\lambda_n$ for temporal links crossing the defect ($\mu=0$, $x_0=L/a-1$, $0\le x_i<L_{\mathrm{d}}/a$), and $K^{(n)}_\mu(x)=1$ elsewhere. This factor is what practically alters the boundary conditions on the defect during the out-of-equilibrium trajectories. Each out-of-equilibrium trajectory (consisting of $\nstep$ steps) is performed every $\nbetween$ standard updates of the OBC system.

The prior distribution is sampled using a 1:4 combination of Cabibbo--Marinari~\cite{Cabibbo:1982zn} heat-bath (HB)~\cite{Creutz:1980zw,Kennedy:1985nu} and over-relaxation (OR)~\cite{Creutz:1987xi} local updating algorithms (from now on, we will refer to this combination simply as ``the standard algorithm''). The topological charge $Q$ is discretized using the customary \emph{clover} discretization, i.e., the simplest CP-odd lattice discretization of $Q$. A naive definition of the lattice topological susceptibility in terms of the clover discretization suffers from multiplicative~\cite{Campostrini:1988cy,Vicari:2008jw} and additive~\cite{DiVecchia:1981aev,DiVecchia:1981hh, Campostrini:1989dh,DElia:2003zne} renormalizations. These are avoided by computing the lattice topological charge on smoothed fields: smoothing suppresses fluctuations below an ultraviolet length scale $R_s$ (the smoothing radius), while leaving the relevant infrared topological fluctuations unaffected (if $R_s$ is taken sufficiently small). Several smoothing algorithms have been proposed in the literature, such as
cooling~\cite{Berg:1981nw,Iwasaki:1983bv,Itoh:1984pr,Teper:1985rb,Ilgenfritz:1985dz,Campostrini:1989dh,Alles:2000sc}, stout smearing~\cite{APE:1987ehd,Morningstar:2003gk}, or gradient flow~\cite{Narayanan:2006rf,Luscher:2009eq,Luscher:2010iy,Luscher:2011bx,Lohmayer:2011si}. When smearing parameters are matched to yield the same $R_s$, different smoothing strategies have all been shown to be numerically equivalent~\cite{Bonati:2014tqa,Alexandrou:2015yba}, and to yield consistent results in the continuum limit (see, e.g., the recent extended critical comparison in Ref.~\cite{Bonanno:2025eeb}). In this study, because of its extremely low computational cost, we adopt Wilson cooling. In this case $R_s = a\sqrt{8 n_{\cool}/3} = \sqrt{8 t}$, where the latter equality allows one to match the number of cooling steps $n_{\cool}$ with the gradient flow time $t$~\cite{Bonati:2014tqa}.

\section{A flow-based approach to topological freezing}
\label{sec:flows}

\begin{figure}
    \centering
    \includegraphics[width=1\linewidth]{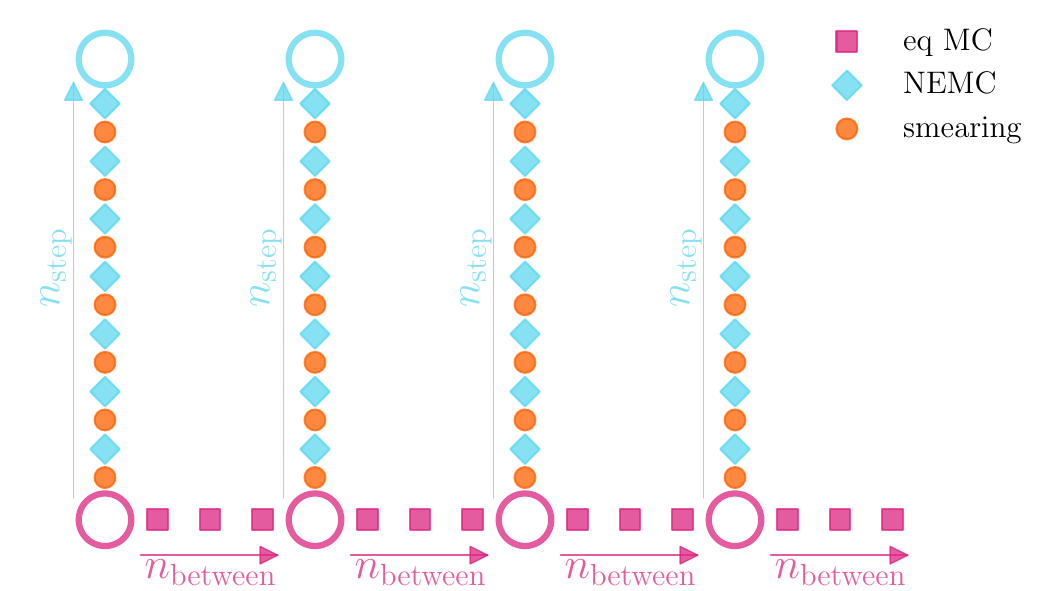}
    \caption{Cartoon of the structure of an SNF simulation: each evolution towards the target distribution alternates non-equilibrium Monte Carlo updates (blue diamonds) with learned stout smearing transformations (orange circles) $\nstep$ times. A new evolution starts every $\nbetween$ standard MCMC updates (pink squares) of a thermalized Monte Carlo chain, the prior.}
    \label{fig:cartoon}
\end{figure}

Our general strategy to mitigate topological freezing is based on two ingredients: the prior, a standard Markov chain simulation performed with an OBC defect of physical size $L_{\mathrm{d}}$ to decorrelate the topological charge, and the flow, which is applied to configurations generated by the prior to remove in an unbiased way any unwanted finite-size effects induced by the OBC defect. This section is dedicated to the description of the features of SNFs, the flow-based approach we chose to guarantee an exact and cheap sampling of the theory with PBC. A schematic representation of an SNF simulation is reported in Fig.~\ref{fig:cartoon}.

\subsection{Stochastic Normalizing Flows for boundary conditions}

NEMC provides an exact framework to transform configurations sampled with OBC into configurations associated with PBC.
Its practical efficiency, however, depends on the dissipation accumulated along the transformation: fast evolutions generally produce broadly distributed reweighting factors, while nearly reversible evolutions may require a large number of Monte Carlo steps. Stochastic Normalizing Flows improve this construction by interleaving the stochastic updates with deterministic, invertible and trainable transformations \cite{wu2020stochastic,Caselle:2022acb,Bulgarelli:2024brv,Bonanno:2025pdp}.

Starting from a configuration $ U^{(0)} $ sampled from the OBC distribution $ q_0 $, an SNF evolution towards the PBC target distribution $p$ is defined by
\begin{equation}
\begin{split}
U^{(0)} \xrightarrow{\,g_{\rho(1)}\,}g_{\rho(1)}(U^{(0)}) \xrightarrow{\,P_{\lambda_1}\,}U^{(1)} & \xrightarrow{\,g_{\rho(2)}\,} \cdots \\
\cdots & \xrightarrow{\,P_{\lambda_{\nstep}}\,}U^{(\nstep)},
\end{split}
\label{eq:snf_evolution}
\end{equation}
where $P_{\lambda_n} $ is a Monte Carlo transition probability satisfying detailed balance with respect to the intermediate action $ S_{\lambda_n} $, while $ g_{\rho(n)} $ is an invertible coupling layer with trainable parameters $ \rho(n) $. The stochastic updates define the annealing protocol between the different boundary conditions, whereas the deterministic layers transport the configurations towards the next intermediate distribution and reduce the dissipation generated by the protocol.

Concerning the Monte Carlo update of $P_{\lambda_n}$, in this work we used the 1:4 combination of HB and OR sweeps,
as earlier described, using the intermediate action $S_{\lambda_{n}}$ defined by a linear protocol for the boundary-condition parameter $\lambda_{n}$.

For what concerns the $ g_{\rho(n)} $ transformations, we employ the defect coupling layers introduced in Refs.~\cite{Bulgarelli:2024yrz,Bonanno:2025pdp}, which act only on links belonging to the defect or on links whose staples contain at least one defect link. Invertibility is ensured through an even--odd and directional masking scheme: during each masked transformation, a set of ``active'' links is updated while the links entering their staples remain frozen. A complete coupling layer is obtained by composing the eight masks associated with the two parities and the four lattice directions. This construction preserves gauge equivariance and yields a tractable Jacobian.

Active links are transformed through gauge-equivariant stout-smearing maps~\cite{Morningstar:2003gk,Nagai:2021bhh,Abbott:2023thq},
\begin{equation}
g_{\rho(n)}\!\left(U_{\mu}(x)\right)=\exp\!\left[i\mathcal{Q}_{\mu}^{(n)}(x;\rho)\right]U_{\mu}(x),
\label{eq:defect_layer}
\end{equation}
where we dropped the superscript $(n)$ on $U_\mu$. The matrix $\mathcal{Q}_{\mu}^{(n)}(x;\rho) $ is Hermitian and traceless and is defined as
\begin{equation}
\begin{split}
\mathcal{Q}_{\mu}^{(n)}(x;\rho)=&\frac{i}{2}\left[\Omega_{\mu}^{(n)\dagger}(x;\rho)-\Omega_{\mu}^{(n)}(x;\rho)\right]- \\&\frac{i}{2N}\operatorname{Tr}\!\left[\Omega_{\mu}^{(n)\dagger}(x;\rho)-\Omega_{\mu}^{(n)}(x;\rho)\right],
\end{split}
\label{eq:defect_Q}
\end{equation}
with
\begin{equation}
\Omega_{\mu}^{(n)}(x;\rho)=C_{\mu}^{(n)}(x;\rho)U_{\mu}^{\dagger}(x),
\label{eq:defect_omega}
\end{equation}
\vspace*{-0.5\baselineskip}
and where $C_{\mu}^{(n)}(x;\rho)$ is equal to
\begin{equation}
\begin{split}
&\sum_{\nu\neq\mu}\left[\rho_{\mu\nu}^{+(n)}(x)U_{\nu}(x)U_{\mu}(x+\hat{\nu})U_{\nu}^{\dagger}(x+\hat{\mu})\,+\right.\\&\left.\rho_{\mu\nu}^{-(n)}(x)U_{\nu}^{\dagger}(x-\hat{\nu})U_{\mu}(x-\hat{\nu})U_{\nu}(x-\hat{\nu}+\hat{\mu})\right].
\end{split}
\label{eq:defect_staples}
\end{equation}

The coefficients $ \rho_{\mu\nu}^{\pm(n)}(x) $ are the trainable parameters of the $n^{\rm{th}}$ coupling layer. Defect coupling layers act only locally around the defect: namely, the coefficients are set to zero unless the transformed active link belongs to the defect or the corresponding staple contains at least one defect link. Links outside the neighborhood of the defect are therefore left unchanged. Since such a deterministic transformation is localized around the defect, its computational cost remains much smaller than that of a global transformation.

\vspace*{-3\baselineskip}
\subsection{Training and sampling}

The path described in Eq.~\eqref{eq:snf_evolution} defines a forward probability distribution
\begin{align}
    \Pf^{\rho} = & \prod_{n=1}^{\nstep} \delta\left(\tilde U^{(n)}, g_{\rho(n)}(U^{(n-1)})\right) P_{\lambda_n} \, \left(\tilde U^{(n)} \rightarrow U^{(n)} \right), \nonumber \\
     =&\prod_{n=1}^{\nstep} P_{\lambda_n} \left(g_{\rho(n)}(U^{(n-1)}) \rightarrow U^{(n)}  \right)
\end{align}
as well as a reverse probability distribution
\begin{align}
    \Pre^\rho = &\prod_{n=1}^{\nstep} P_{\lambda_n} (U^{(n)} \rightarrow \tilde U^{(n)}) \, \delta\left(U^{(n-1)}, g^{-1}_{\rho(n)}(\tilde U^{(n)})\right), \nonumber \\
    =&\prod_{n=1}^{\nstep} J_{g_{\rho(n)}}(U^{(n-1)}) \, P_{\lambda_n} (U^{(n)} \rightarrow g_{\rho(n)}(U^{(n-1)})).
\end{align}
where we used the change of variables theorem in the last step.
Here $S_{\lambda_{n}}$ is the intermediate action defined by the boundary condition parameter $\lambda_n$ being changed along the evolution, while $ J_{g_{\rho(n)}} $ is the Jacobian determinant of the corresponding coupling layer. 

With these definitions it is possible to derive Crooks' fluctuation theorem for MCMC~\cite{Crooks_1999}:
\begin{equation}
    \frac{p(U^{(\nstep)}) \Pre^\rho}{q_0(U^{(0)})\Pf^{\rho}} = e^{-(W_\rho - \Delta F)},
\end{equation}
where one finds the crucial quantity to be computed along the forward SNF evolutions of Eq.~\eqref{eq:snf_evolution}, i.e., the generalized work
\begin{equation}
\label{eq:snf_generalized_work}
\begin{split}
W_{\rho}=\sum_{n=0}^{\nstep-1}\{&S_{\lambda_{n+1}}\!\left[g_{\rho(n+1)}(U^{(n)})\right]+\\
&-S_{\lambda_n}(U^{(n)})-\log J_{g_{\rho(n+1)}}(U^{(n)})\}.
\end{split}
\end{equation}
Note that for the evolution from OBC to PBC considered here, the change in the action is localized around the three-dimensional defect. 

The generalized work is essential, as it defines the reweighting factor needed for an unbiased sampling of the target probability distribution. In particular, using Crooks' therome one can write expectation values of observables in a system with PBC as
\begin{equation}
\langle\mathcal{O}\rangle_{p}=\frac{\left\langle\mathcal{O}(U^{(\nstep)})\,\mathrm{e}^{-W_{\rho}}\right\rangle_{\mathrm{f}}}{\left\langle \mathrm{e}^{-W_{\rho}}\right\rangle_{\mathrm{f}}},
\label{eq:snf_reweighting}
\end{equation}
where the average is taken over forward (OBC $\to$ PBC) SNF evolutions, denoted by ``f''.\footnote{While not of practical use here, SNFs provide a way to compute the difference in free energy $ \Delta F = F_{\rm PBC} - F_{\rm OBC}$ through Jarzynski equality $\left\langle \mathrm{e}^{-W_{\rho}}\right\rangle_{\mathrm{f}} = \mathrm{e}^{-\Delta F}$.} The observable is calculated at the last step, see Eq.~\eqref{eq:snf_evolution}.

Furthermore, the generalized work can be directly related to the reverse Kullback--Leibler (KL) divergence between the forward and reverse path distributions:
\begin{equation}
\DKL \!\left(q_0 \Pf^{\rho}\Vert p \Pre^{\rho}\right)=\left\langle W_{\rho}\right\rangle_{\mathrm{f}}-\Delta F\geq 0.
\label{eq:snf_kl}
\end{equation}
This is equal to the average of the so-called ``dissipated work'' $\neavg{\Wd} = \neavg{W_\rho} - \Delta F$. The inequality comes directly from the properties of KL divergences and it is just a statement of the Second Law of Thermodynamics: the smaller the dissipation, the more reversible SNF evolutions become.

A more quantitative metric to determine the efficiency of the reweighting estimator of Eq.~\eqref{eq:snf_reweighting} is the Effective Sample Size (ESS). It is formally defined as $\ESS = \Var(\mathcal{O})/\Var_{\mathrm{f}}(\mathcal{O})$: the numerator is the theoretical variance of the observable $\mathcal{O}$ while the denominator is the variance of the estimator of Eq.~\eqref{eq:snf_reweighting}. In practice, we use an observable-agnostic different estimator, simply defined as 
\begin{equation}
\label{eq:ess}
    \hESS = \frac{\langle \exp(-W_\rho) \rangle_{\mathrm{f}}^2}{\langle \exp(-2W_\rho)\rangle_{\mathrm{f}}} = \frac{1}{\langle \exp(-2\Wd) \rangle_{\mathrm{f}}}.
\end{equation}
This quantity takes values between $0$, which signals that the reweighting procedure is infinitely noisy, and $1$, indicating a perfectly efficient estimator.

Let us look at the training procedure. Since the free-energy difference $ \Delta F $ is independent of $ \rho $, the coupling layers can be trained by directly minimizing $ \langle W_{\rho}\rangle_{\mathrm{f}} $. This makes the transformation more reversible, reduces the fluctuations of the non-equilibrium weights and increases the ESS. Furthermore, we train each coupling layer separately: more precisely, we minimize each of the $\nstep$ terms in Eq.~\eqref{eq:snf_generalized_work} independently, following Refs.~\cite{Bulgarelli:2024brv,Bonanno:2025pdp}. This procedure already highlights the main feature of training SNFs: the protocol dictates in an explicit way the path between prior and target; what is left is to optimize the efficiency of the individual steps.

The only $\rho^{\pm(n)}_{\mu \nu}(x)$ parameters that are non-zero and that are changed in the training procedure are either 1) used in a smearing transformation of a link on the defect or 2) associated with a staple containing at least one link on the defect. It is possible to classify these parameters in a geometric fashion: in this work we identify 9 approximate classes,  following exactly the same procedure defined in Ref.~\cite{Bonanno:2025pdp} (we point the interested reader to Appendix A therein for more details). In practice, we perform the training only for $\nstep^{(\rm train)}=16$, average for each step the parameters belonging to the same class and interpolate each of the 9 classes over all steps with a spline function. This interpolation function is the actual result of the training and can be used to generate the $\rho^\pm_{\mu \nu}(x; n)$ for any value of $n \in [1, \nstep]$ (see Table~\ref{tab:snf_parameters}); $\nstep^{(\rm train)}=16$ is large enough to resolve the behavior of each class of parameters along the protocol. 

The training is done separately for each value of $\beta$ and $L_{\mathrm{d}}/a$ for about 1000--2000 gradient descent steps, always on an $(L^{(\rm train)}/a)^4 = 16^4$ lattice. We consistently obtain SNFs which are approximately 3 times as efficient as a standard NEMC. We then directly transfer the generated $\rho^{\pm(n)}_{\mu \nu}(x)$ to the large lattices of Table~\ref{tab:simulation_parameters}, obtaining the same improvement factor; this confirms that the defect coupling layers act very locally around the defect and, at this level of expressiveness, are unaffected by global properties of the lattice. 

The training procedure we just described entails negligible costs with respect to the sampling effort: this is clear when comparing $L^{(\rm train)}/a$ and $\nstep^{(\rm train)}$ with the values used for the simulations, see Tables~\ref{tab:simulation_parameters} and \ref{tab:snf_parameters}. We remark that this is linked to the relative simplicity of the stout-smearing transformations at the heart of the defect coupling layers. Architectures that include larger loops or deeper smearing levels (i.e., encoding more physical content) might transfer less effectively to larger volumes and be less amenable to interpolation along SNF evolutions.

\vspace*{-\baselineskip}
\section{Numerical results}
\label{sec:results}

\vspace*{-0.5\baselineskip}
\subsection{Scaling of costs in the continuum}
\vspace*{-0.5\baselineskip}

We measure the effectiveness of our SNF at sampling a given observable $\mathcal{O}$ according to the target probability distribution using the variance of the estimator times the cost of obtaining one sample in units of Monte Carlo updates. In practice, for the sampling of topological observables the effective cost is formally defined as
\begin{equation}
    \Var_{\rm f}(Q^2) \, 2\tauint(Q^2) \, \left(\nstep + \nbetween\right)
\end{equation}
where $\Var_{\rm f}$ accounts for the larger variance due to the presence of the reweighting procedure, $\tauint$ for any residual autocorrelation in the final samples, $\nstep$ for the cost of a single SNF evolution and $\nbetween$ for the number of equilibrium updates between subsequent out-of-equilibrium evolutions. Approximating the true effective sample size with $\hESS$, we obtain the effective cost we measure in practice in our simulations:
\begin{equation}
\label{eq:benchmark}
 \mathrm{cost} = \Var(Q^2) \, \frac{2\tauint(Q^2)}{\hESS} \, \left(\nstep + \nbetween\right).
\end{equation}
We wish to control this benchmark metric in the continuum limit by keeping the ratio $\tauint(Q^2)/\hESS$ constant while calibrating $\nstep$ and $\nbetween$ accordingly.

Our strategy begins by fixing the size of the OBC defect in the ballpark of $0.4 \sqrt{8 t_0}$, see Tab.~\ref{tab:snf_parameters}, for the runs whose parameters are reported in Tab.~\ref{tab:simulation_parameters}. The chosen defect size has been shown in previous studies employing Parallel Tempering on Boundary Conditions~\cite{Bonanno:2024nba,Bonanno:2025eeb} to be a large enough value to allow for an efficient decorrelation of the topological charge in the system with OBC. Still, $\tauint(Q^2)$ is expected to grow in the continuum limit, in good approximation, as $a^{-2}$~\cite{Luscher:2011kk,McGlynn:2014bxa}: to counter this, we increase the spacing between different evolutions exactly as $\nbetween \sim a^{-2}$, see Tab.~\ref{tab:snf_parameters}. 

\begin{table}[!t]
    \centering
    \begin{tabular}{c|c|c|c|c|c|c}
    $\beta$ & $a/\sqrt{8t_0}$ & $L/a$ & $L / \sqrt{8t_0}$ & $L_{\mathrm{d}}/a$ & $L_{\mathrm{d}}/\sqrt{8 t_0}$ & $n_{\cool}$ max\\
    \hline
    $6.4$ & $0.1086$ & $30$ & $3.26$ & $3$ & $0.326$ & 60\\
    $6.5$ & $0.0945$ & $34$ & $3.21$ & $4$ & $0.378$ & 60\\
    $6.6$ & $0.0828$ & $38$ & $3.15$ & $5$ & $0.414$ & 60 \\
    $6.7$ & $0.0726$ & $44$ & $3.19$ & $6$ & $0.436$ & 60\\
    $6.8$ & $0.0638$ & $50$ & $3.19$ & $6$ & $0.383$ & 80\\
    \end{tabular}
    \caption{Lattice spacing $a$, lattice size $L$ and defect size $L_{\mathrm{d}}$ in units of both $a$ and the scale $\sqrt{8t_0}$, for all the ensembles used in this work. The determinations of $t_0$ come from interpolations of the data of Ref.~\cite{Giusti:2018cmp}. We also report the maximum number of cooling steps performed for each run.}
    \label{tab:simulation_parameters}
\end{table}

\begin{table}[!t]
    \centering
    \begin{tabular}{c|c|c|c|c|c|c}
    $\beta$ & $L_{\mathrm{d}}/a$ & $\nstep$ & $\nbetween$ & Statistics & $\tauint(Q^2)$ & $\hESS$ \\
    \hline
    $6.4$ & $3$ & $200$  & $100$ &  4.4k & 1.3(2) & 0.403(19) \\
    $6.5$ & $4$ & $475$  & $130$ & 12.0k & 2.1(2) & 0.401(9)  \\
    $6.6$ & $5$ & $975$  & $200$ & 11.8k & 1.9(2) & 0.363(13) \\
    $6.7$ & $6$ & $1600$ & $260$ & 10.7k & 2.1(2) & 0.337(12) \\
    $6.8$ & $6$ & $1600$ & $340$ &  4.3k & 3.6(7) & 0.351(16) \\
    \end{tabular}
    \caption{Parameters for SNF evolutions for all ensembles. The column ``Statistics'' refers to the number of collected samples of the topological charge. The analysis was performed using the $\Gamma$-method~\cite{Wolff:2003sm} with the pyerrors package~\cite{Joswig:2022qfe}.}
    \label{tab:snf_parameters}
\end{table}

\begin{figure}[t]
    \centering
    \includegraphics[width=0.95\linewidth]{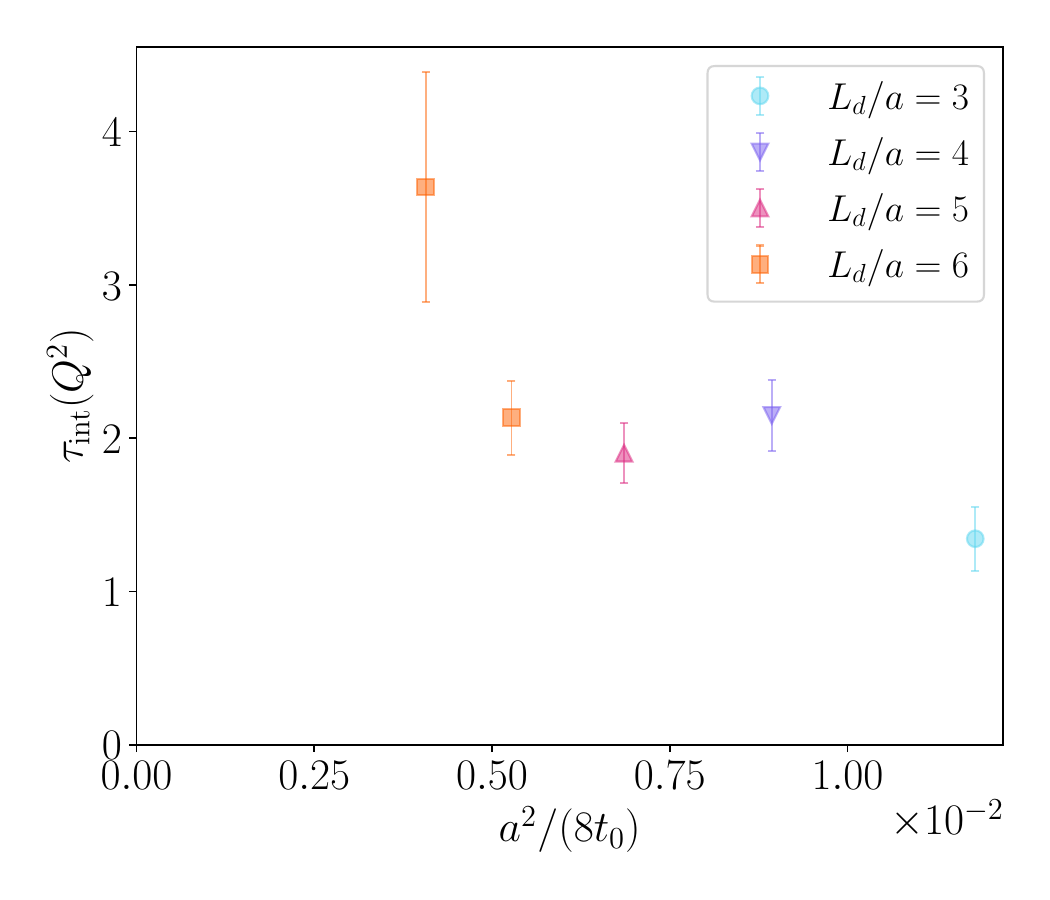}
    \caption{Integrated autocorrelation time of $Q^2$ for the flows described in Tab.~\ref{tab:snf_parameters} for varying lattice spacings.}
    \label{fig:tau_a}
\end{figure}

Fig.~\ref{fig:tau_a} shows that integrated autocorrelation times for $Q^2$ are remarkably constant when following the strategy outlined above. The only exception may be the finest lattice, for which the limited number of samples does not allow for a determination of $\tauint(Q^2)$ as accurate as those for the coarser points. We suspect that a non-trivial dependence on $\nstep$ could explain the mismatch, as further decorrelation may happen with mixed boundary conditions along the evolution itself. We leave a more quantitative study of this phenomenon to future work. In any case, the obtained small values of $\tauint(Q^2)$ show that SNF allow for an ergodic sampling of the topological charge down to lattice spacings as fine as $\sim 0.03$ fm (we assume $\sqrt{8t_0}=0.475$ fm~\cite{Giusti:2018cmp}). To further support this claim, we also provide the histograms of the topological charge at the end of the out-of-equilibrium evolution, when PBC are restored. An example for $\beta=6.7$ is shown in Fig.~\ref{fig:Q_histo_b6.7}, other plots can be found in the Appendix.

\begin{figure}[!t]
\centering
\includegraphics[scale=0.48]{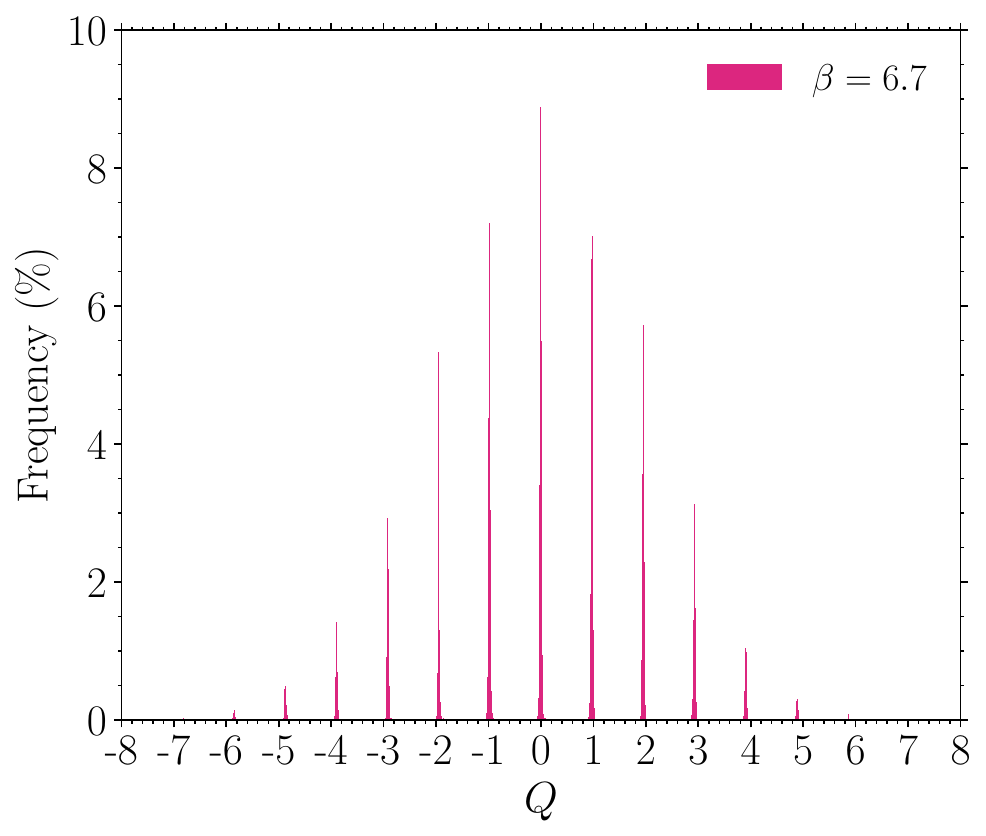}
\caption{Example of histogram of the topological charge at the end of the out-of-equilibrium evolutions with SNF.}
\label{fig:Q_histo_b6.7}
\end{figure}

Next, we look at the relationship between $\hESS$ and $\nstep$. The scaling of $\nstep$ with respect to the number of the relevant degrees of freedom is a crucial feature of NEMC and it is retained by SNFs as well. It has been numerically tested in a variety of models~\cite{Bonanno:2024udh,Bulgarelli:2024brv,Bulgarelli:2025riv}, showing that metrics such as the dissipated work or the $\hESS$ are actually simple functions of $\nstep$ and $\ndof$, respectively the duration of the non-equilibrium evolution in units of Monte Carlo updates and the number of degrees of freedom affected by the change in the action throughout a given protocol.

In particular, for evolutions in the boundary conditions it has been observed that the only relevant variable is $\nstep/\ndof$: in other words, to keep the aforementioned metrics constant (i.e., to guarantee in good approximation the same efficiency when sampling with this family of flows) the number of non-equilibrium Monte Carlo updates must grow with the size of the defect in lattice units. In practice, we wish to keep the $\hESS$ constant with
\begin{equation}
    \nstep \propto\left(\frac{L_{\mathrm{d}}}{a}\right)^3 \sim a^{-3}
\end{equation}
where we made explicit the scaling in the continuum limit when keeping the defect size $L_{\mathrm{d}}$ fixed in physical units.
This scaling has been numerically checked in the $\SU(3)$ pure gauge theory in $3+1$ dimensions both at coarse and intermediate lattice spacings, see Ref.~\cite{Bonanno:2025pdp}. 

\begin{figure}[t]
    \centering
    \includegraphics[width=\linewidth]{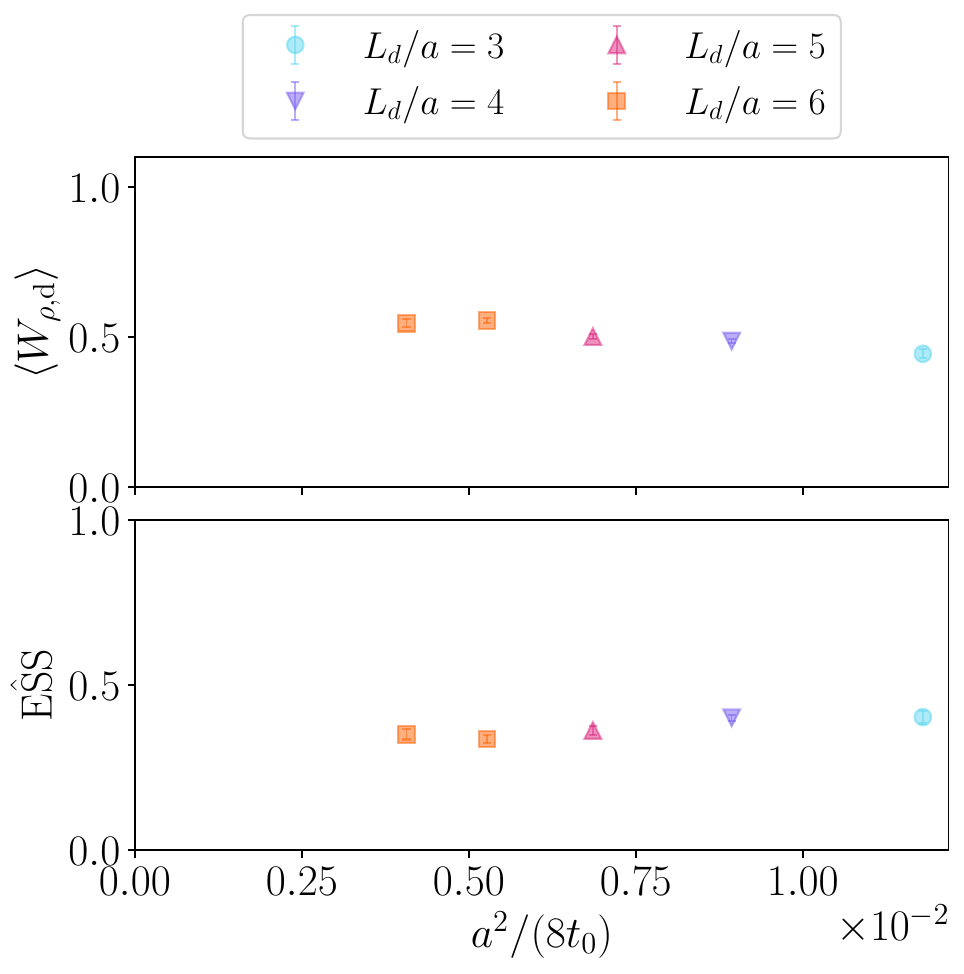}
    \caption{Average dissipated work (upper panel) and ESS (lower panel) for the flows described in Tab.~\ref{tab:snf_parameters} for varying lattice spacings.}
    \label{fig:dkless_a}
\end{figure}

We follow exactly this recipe and rescale $\nstep$ accordingly in the continuum limit, see Tab.~\ref{tab:snf_parameters}: in Fig.~\ref{fig:dkless_a} it is possible to appreciate how both the dissipated work and the $\hESS$ are kept remarkably constant by an appropriate increase in the length of SNF evolutions. A weak but visible trend is also noticeable both in the dissipation (slightly growing as $a$ decreases) and in the ESS (slightly decreasing as $a$ decreases): this behavior is compatible with the fact that the coupling $\lambda_n \beta$ on the defect has to reach larger and larger values when approaching the continuum limit. In fact, this scaling was already examined in detail in Ref.~\cite{Bulgarelli:2024brv} when studying flows in $\beta$; for the sake of simplicity, we chose not to rescale $\nstep$ accordingly. Our results show that this is indeed a visible but subleading effect.

\begin{figure}[t]
    \centering
    \includegraphics[width=\linewidth]{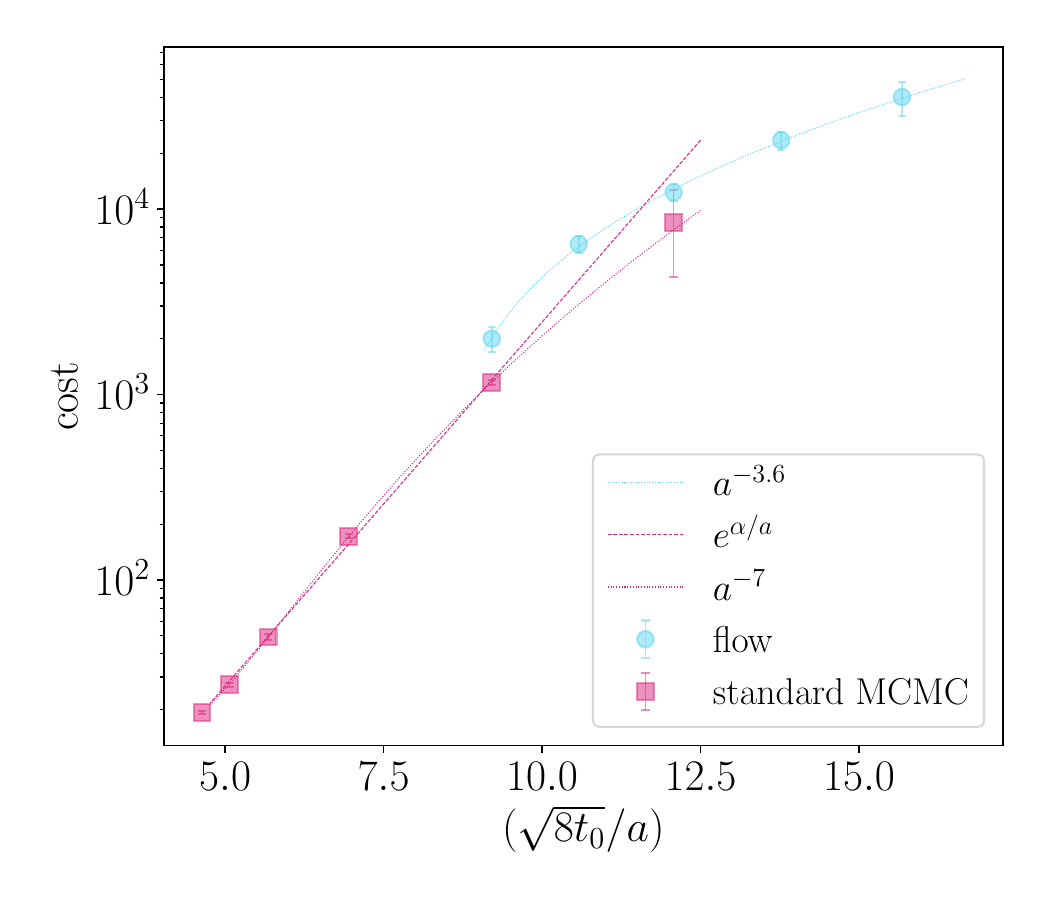}
    \caption{Benchmark comparison between the flows described in Tab.~\ref{tab:snf_parameters} and standard MCMC simulations, cf.~Tab.~\ref{tab:tauint_std}. We use $2\tauint(Q^2)/\hESS \, (\nstep + \nbetween)$ as the cost function for SNFs and $2\tauint(Q^2) \nbetween$ for the standard MCMC.}
    \label{fig:comparison}
\end{figure}

We are finally in a position to check the efficiency of our method against a standard MCMC simulation. To do so, we use the benchmark defined in Eq.~\eqref{eq:benchmark}, ignoring the $\Var(Q^2)$ factor (which ought to be the same for both algorithms). This is a fair comparison, as this benchmark is measured in units of the same Monte Carlo update (1HB+4OR): for a standard MCMC, we naturally have $\ESS=1$ and $\nstep=0$, whereas $\nbetween$ takes the role of the spacing between each measurement on a sequential chain. In practice, in Fig.~\ref{fig:comparison} we compare $2\tauint(Q^2)/\hESS \, (\nstep + \nbetween)$ for SNFs with $2\tauint(Q^2) \, \nbetween$, using measurements of this quantity obtained with the standard algorithm, reported in Tab.~\ref{tab:tauint_std}.

\begin{table}[!t]
    \centering
    \begin{tabular}{c|c|c|c|c|c|c}
    $\beta$ & $L/a$ & $a/\sqrt{8t_0}$ & $L/\sqrt{8t_0}$ & $\nbetween$ & Statistics & $\tauint(Q^2)$\\
    \hline
    $5.95$ & 16 & 0.2157 & 3.45 & $10$  & 1.94M & 0.980(40) \\
    $6.00$ & 16 & 0.1974 & 3.16 & $10$  & 2.03M & 1.40(50)  \\
    $6.07$ & 18 & 0.1761 & 3.17 & $10$  & 1.94M & 2.50(10)  \\
    $6.20$ & 22 & 0.1439 & 3.17 & $10$  & 1.23M & 8.50(50)  \\
    $6.40$ & 30 & 0.1086 & 3.26 & $100$ & 0.71M & 5.75(30)  \\
    $6.60$ & 38 & 0.0828 & 3.15 & $20$  & 0.02M & 212(104) \\
    \end{tabular}
    \caption{Autocorrelation times obtained for the standard algorithm (1HB+4OR) in units of measurements, taken every $\nbetween$ updating steps. Data up to $\beta=6.4$ come from~\cite{Bonanno:2025eeb}; the value at $\beta=6.6$ was determined in this study and represents only a rough estimate due to the limited statistics.}
    \label{tab:tauint_std}
\end{table}

The results for the standard MCMC are compatible both with an exponential law $\tauint\sim\mathrm{e}^{\alpha/a}$ and with a power law with very large exponent, $\tauint\sim a^{-7}$. For the flow-based approach, instead, the benchmark follows a power law with a much smaller exponent of about $3.6$, very similar to the behavior of $a^{-3} + a^{-2}$ we expect from the rescaling of $\nstep$ and $\nbetween$ in Eq.~\eqref{eq:benchmark}. While a more rigorous study is needed to test the fine details of the behavior in the continuum limit of both approaches, such a result is a clear indication that the scaling of the costs to sample topological observables in the continuum limit is effectively mitigated by our SNF-based approach. 

At this point a natural question arises: when are SNFs effectively cheaper than standard MCMC? Looking again at Fig.~\ref{fig:comparison}, our flow-based approach is certainly more expensive up to $a /\sqrt{8t_0} \sim 0.1$ ($a \sim 0.048$fm), but it is fair to argue that it becomes more efficient below $a /\sqrt{8t_0} \sim 0.08$ ($a \sim 0.038$fm). 
It is important to stress, however, that a fair comparison is not simply a matter of efficiency, but whether it is even possible to give a reliable estimate of the error. Indeed, the very determination of $\tauint(Q^2)$ for a standard MCMC is a difficult process, which requires extremely large statistics and a careful analysis just to capture the correct behavior of the autocorrelation function at large Monte Carlo time separation~\cite{Schaefer:2010hu}: the huge error at $\beta=6.6$ is a clear signal of this.

Furthermore, the main source of the computational costs for our approach is essentially due to $\nstep$, which controls the \emph{coefficients} of the scaling law presented in Fig.~\ref{fig:comparison}. Improving the defect coupling layer structure in order to further decrease the dissipation during SNF evolutions would push the scaling curve further down, making it much more affordable while preserving its scaling with $a$.

\vspace*{-0.5\baselineskip}
\subsection{Topological susceptibility at fine spacings}
A determination of the topological susceptibility $\chi$ can be extracted from our fine-lattice-spacing runs (all raw numerical data can be found in Appendix~\ref{app:numerical_data}). Based on the computational resources at our disposal, we expected to accumulate statistics large enough to achieve a $\sim 5\%$ uncertainty on $\chi$ across our ensembles (about 10k samples each). Our results for $\beta=6.40,6.50,6.60,6.70$ feature a $\sim 4.5\%$ relative error, showing excellent control of the scaling of the sampling method in a realistic setting. The only exception is the $\sim 9\%$ uncertainty for $\beta=6.80$: this is due both to more limited statistics for this ensemble (about 4k samples) and to the larger residual autocorrelations that are visible in Fig.~\ref{fig:tau_a}.

These determinations can be compared with the continuum extrapolation of the high-statistics (order $10^6$ samples) data at coarser lattice spacings $\beta=5.95,6.00,6.07,6.20,6.40$ obtained for $\chi$ in the $\SU(3)$ Yang--Mills theory in Ref.~\cite{Bonanno:2025eeb}, which have a precision of about $0.3\%-0.6\%$. Such a comparison suffices for the main goal of the present paper, which is not to provide a novel precise determination of $\chi$, but rather to prove that topology autocorrelations (and thus uncertainties on the susceptibility) are completely under control. In the following, all data refer to a fixed cooling smoothing radius of $R_s/\sqrt{8t_0}\simeq 0.918$, which corresponds to an equivalent gradient flow time $t/t_0\simeq0.843$. To obtain data at the same $R_s/\sqrt{8t_0}$, we performed a cubic spline interpolation of $a^4\chi$ as a function of $n_{\cool}$.

\begin{figure}[!t]
    \centering
    \includegraphics[scale=0.5]{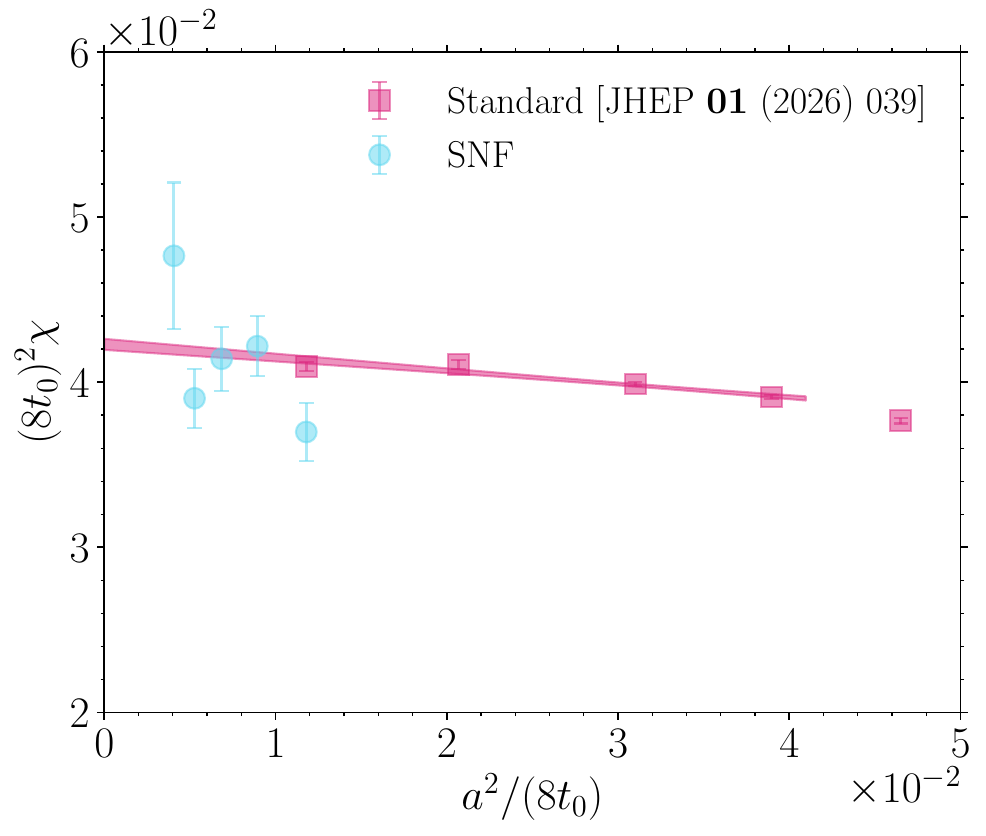}
    \caption{Comparison of the new SNF data for the topological susceptibility $\chi$ on fine lattice spacings with the continuum extrapolation of data obtained for coarser lattice spacings and with the standard algorithm in~\cite{Bonanno:2025eeb}.}
    \label{fig:cont_lim_chi}
\end{figure}

SNF data for the topological susceptibility are shown in Fig.~\ref{fig:cont_lim_chi}: as can be seen, the new results at fine lattice spacings agree (within their larger error bars) with the continuum extrapolation of coarser determinations obtained with the standard algorithm and fully periodic boundaries (all falling in a regime where freezing is not yet significant). In particular, the $\beta=6.40,6.50,6.60,6.70,6.80$ SNF data for $\chi$ deviate by at most 2.5 standard deviations from the continuum extrapolation curve (described by means of $\mathcal{O}(a^2)$ leading lattice artifacts). Let us also comment on the fact that the continuum limit obtained in~\cite{Bonanno:2025eeb}, $(8 t_0)^2\chi = 0.04229(33)$, perfectly agrees with other recent determinations in the literature~\cite{Ce:2015qha,Bonati:2015sqt,Durr:2025qtq}. Just as an example, Ref.~\cite{Ce:2015qha} reports $(8t_0)^2 \chi = 0.04269(45)$ (see~\cite{Bonanno:2025eeb} for a more extended comparison), agreeing within 0.7 standard deviations. Thus, the SNF results obtained for the topological susceptibility on fine lattice spacings are in good agreement with previous continuum determinations too.

\section{Conclusions}
\label{sec:conclusions}

In this work we have provided robust numerical evidence for a computational strategy that, when approaching the continuum limit, displays a growth of computational costs for topological observables which is much milder than the standard MCMC approach. This was obtained by combining open boundary conditions and Stochastic Normalizing Flows: the former decorrelate topological modes, with residual autocorrelation times that scale approximately as $a^{-2}$, while the latter are used for the exact removal of boundary effects, the cost of which scales approximately as $a^{-3}$. The main result of this paper is the ability to compute the real costs and their scaling, and compare them fairly with the MCMC baseline.

We want to stress that, in a direct comparison with standard local updates, this architecture is still expensive in absolute terms up to relatively fine lattice spacings: the scaling with $1/a$ is favorable, but its \emph{coefficients} are still large. The next step in this line of research will indeed be the establishment of a systematic improvement program directly targeted at reducing these coefficients, while maintaining the same scaling behavior. This program is articulated in two main directions: the development of a protocol for the boundary-condition parameter $\lambda_n$ that optimizes the dissipation for a given $\nstep$, and the integration of more complicated and expressive defect coupling layers that are able to remove dissipation in a more efficient way.

For what concerns the future of state-of-the-art techniques in sampling gauge theories in the continuum limit, we must necessarily mention PTBC~\cite{Hasenbusch:2017unr,Bonanno:2020hht,Bonanno:2024zyn,Bonanno:2025eeb}. This approach has proven to be a highly efficient sampler of topological observables and displays a favorable scaling for $a\to 0$, see, e.g., Refs.~\cite{Bonanno:2024zyn,Bonanno:2025eeb}. While a quantitative comparison of the costs between PTBC and our method at very fine lattice spacings would be quite useful, it is beyond the scope of this work; furthermore, it is made difficult by the different requirements in terms of memory of the two approaches. Instead, we look at this from a different perspective, as the \emph{combination} of NEMC/SNFs with PTBC is a truly exciting prospect. More specifically, flow-based approaches could significantly boost the performance of PTBC by reducing the number of independent replicas that must be concurrently simulated, while at the same time preserving its favorable scaling towards the continuum limit (for other recent algorithmic proposals leveraging NFs to boost parallel tempering schemes see, e.g., Refs.~\cite{Invernizzi:2022aaa,Abbott:2024mix,Zhang:2025aaa}). We are planning to pursue this approach in the near future.

Furthermore, the method presented in this work could be combined with recent developments in lattice field theory that leverage deep generative models to construct hierarchical multiscale samplers~\cite{Abbott:2024knk,Bauer:2024byr,Singha:2025lsd,Singha:2026slc,Hasenfratz:2026azk,Hasenfratz:2026mae,Singha:2026qyx}.
While our approach primarily targets topological freezing, these multiscale methods aim to mitigate the more conventional critical slowing down associated with long-distance modes and increasing lattice volumes. Their combination could therefore provide a promising strategy for simultaneously addressing both sources of critical slowing down.

Finally, we envision the application of the NEMC/SNF family of architectures to other physically interesting setups. In particular, we would like to mention the study of first-order phase transitions: on the one hand, standard MCMC sampling suffers from a form of supercritical slowing down~\cite{Berg:1991cf}, so an efficient sampler is sorely needed (for recent algorithmic strategies employed for the mitigation of supercritical slowing down close to a first-order point, see, e.g., Refs.~\cite{Langfeld:2012ah,Borsanyi:2022xml,Lucini:2023irm,Bennett:2025whm,Giusti:2025fxu,Wenger:2026sjp}); on the other hand, such transitions encapsulate valuable physical information in observables such as the latent heat or the interface tension, which are hard to measure at high precision with standard methods.

\begin{acknowledgments}
\noindent A.~N., D.~P. and L.~V. acknowledge support by the Simons Foundation grant 994300 (Simons Collaboration on Confinement and QCD Strings). E.~C. acknowledges support by the European Union (ERC, FoundLatt, 101212401). Views and opinions expressed are however those of the author only and do not necessarily reflect those of the European Union or the European Research Council. Neither the European Union nor the granting authority can be held responsible for them. A.~N. acknowledges support from the European Union -- Next Generation EU, Mission 4 Component 1, CUP D53D23002970006, under the Italian PRIN ``Progetti di Ricerca di Rilevante Interesse Nazionale -- Bando 2022'' prot. 2022ZTPK4E. A.~N., D.~P. and L.~V. acknowledge support from the SFT Scientific Initiative of INFN. The work of D.~V.~is supported by STFC under Consolidated Grant No.~ST/X000680/1. A.~B. was supported by the Deutsche Forschungsgemeinschaft (DFG, German Research Foundation) as part of the CRC 1639 NuMeriQS -- project no.\ 511713970 and under Germany's Excellence Strategy -- Cluster of Excellence ``Color meets Flavor'' (CmF) EXC 3107 -- 533766364. We acknowledge the EuroHPC Joint Undertaking for awarding the project EHPC-DEV-2024D11-010 access to the EuroHPC supercomputer LEONARDO hosted by CINECA, Italy through an EuroHPC Development Access call. We also acknowledge ISCRA for awarding the Class B project TSOOEE - HP10B7O173 to the LEONARDO supercomputer, owned by the EuroHPC Joint Undertaking, hosted by CINECA (Italy). This work was partially carried out using the computational facilities of the ``Lovelace'' \href{https://www.plymouth.ac.uk/about-us/university-structure/faculties/science-engineering/hpc}{High Performance Computing Centre, University of Plymouth}.
\end{acknowledgments}

\section*{Data Availability}
\noindent Data supporting the findings of this article are available in the main text and in the Appendix. Further data available upon reasonable request.

\renewcommand*{\bibfont}{\small}
\bibliographystyle{apsrev4-2}
\bibliography{biblio}

%\clearpage
\onecolumngrid
\appendix

\setcounter{equation}{0}
\renewcommand{\theequation}{A.\arabic{equation}}

\setcounter{table}{0}
\renewcommand{\thetable}{A.\Roman{table}}

\setcounter{figure}{0}
\renewcommand\thefigure{A.\arabic{figure}}   

\section{Raw numerical data and additional plots}
\label{app:numerical_data}

\begin{figure}[!htb]
\centering
\includegraphics[scale=0.48]{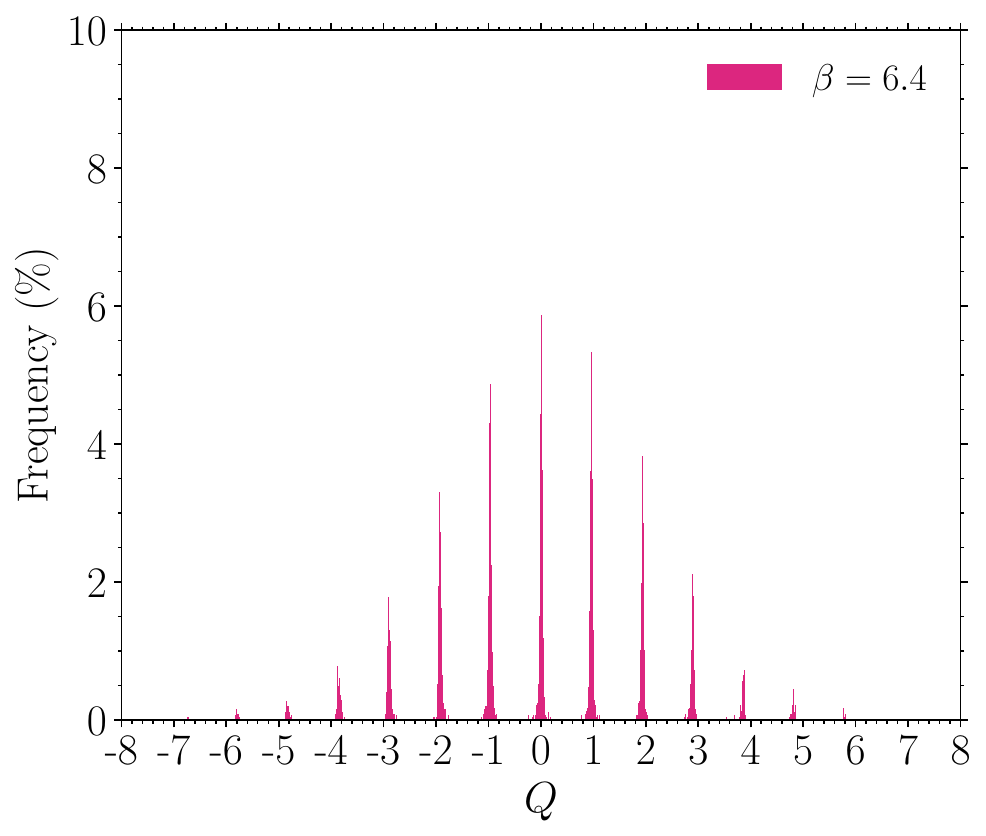}
\includegraphics[scale=0.48]{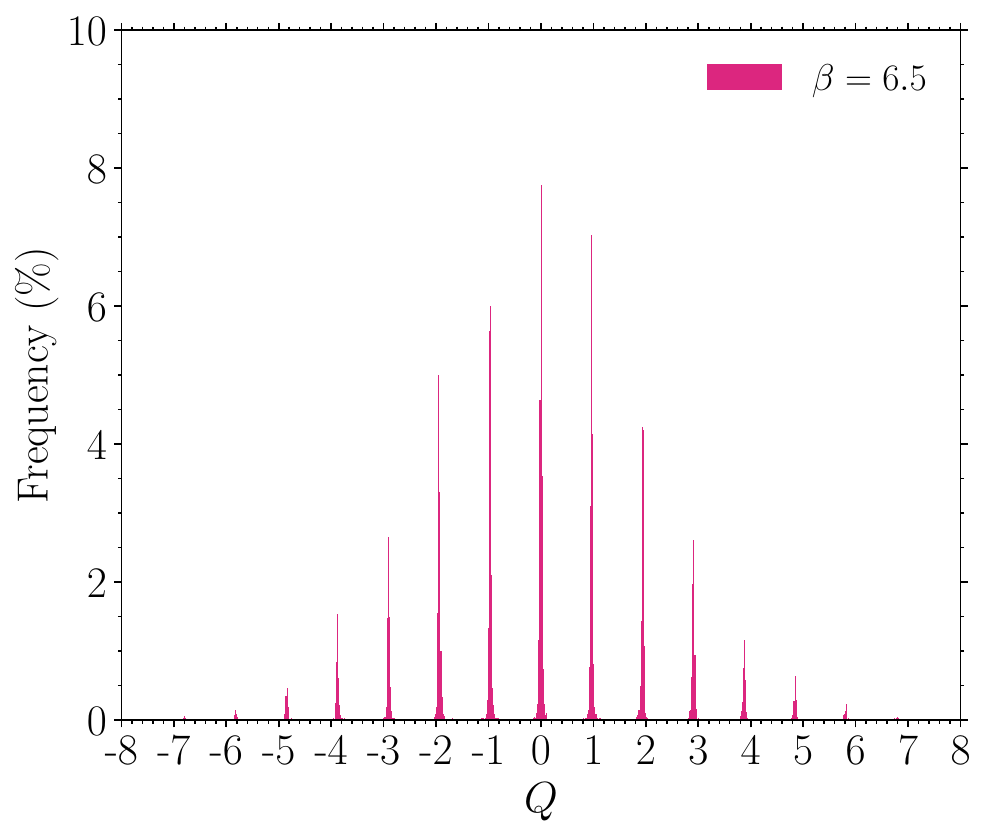}\\
\includegraphics[scale=0.48]{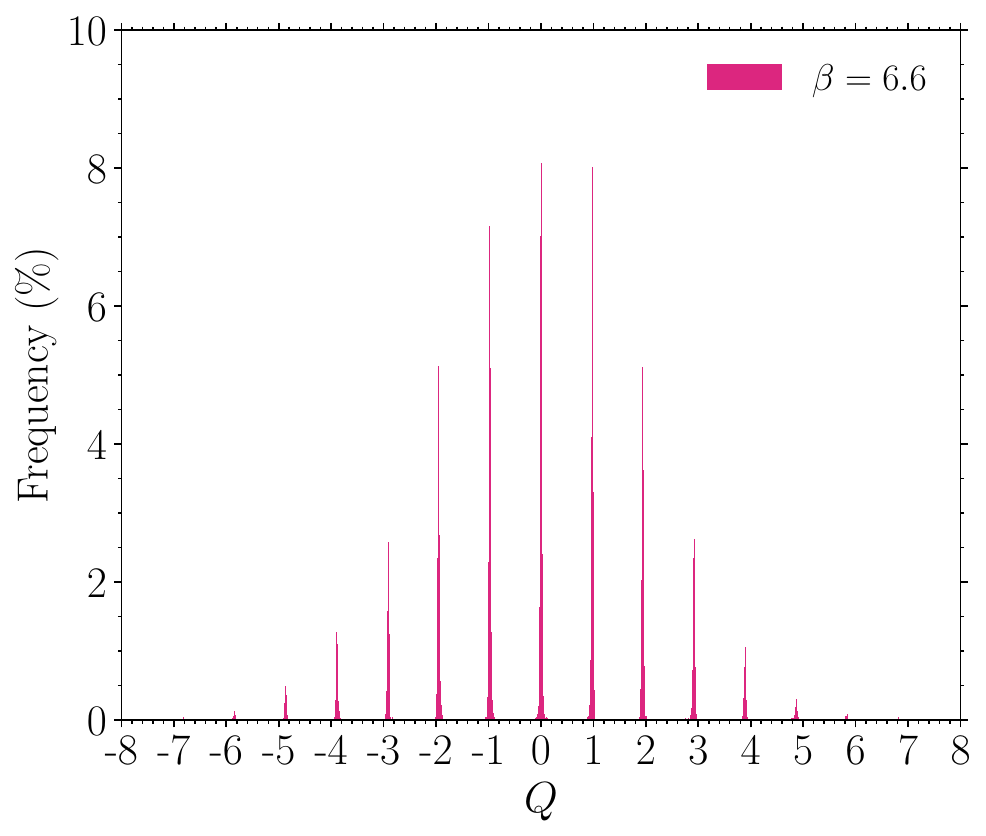}
\includegraphics[scale=0.48]{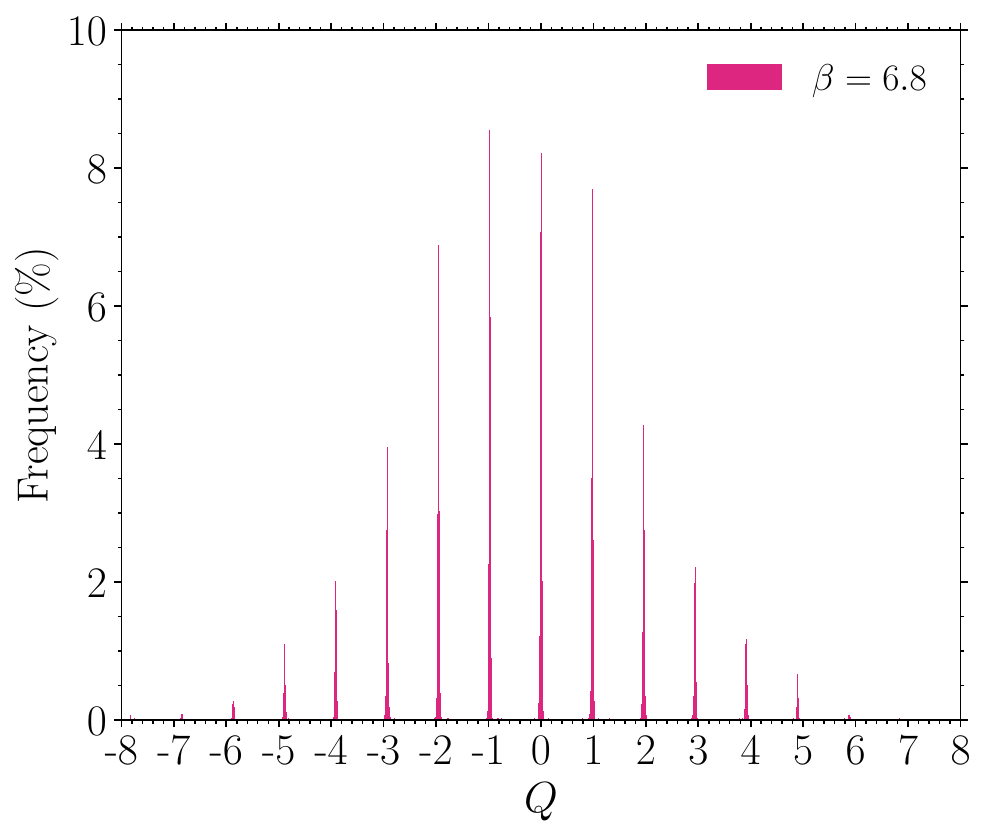}
\caption{Further histograms of the topological charge at the end of the out-of-equilibrium evolutions with SNF.}
\label{fig:Q_histo_other_b}
\end{figure}

This appendix collects all raw numerical data for the topological susceptibility as a function of $n_{\cool}$, as well as the topological susceptibility data in physical units appearing in Fig.~\ref{fig:cont_lim_chi}. Raw numerical data for the runs with the standard algorithm can be found in Ref.~\cite{Bonanno:2025eeb}. Moreover, in Fig.~\ref{fig:Q_histo_other_b}, we show other histograms of the topological charge samples (computed at the end of the out-of-equilibrium trajectory, where PBC are restored) from SNF runs.

\begin{table}[!htb]
\centering
\begin{minipage}{0.49\textwidth}
\centering
\footnotesize
\begin{tabular}{cc|cc|cc|cc|cc}
\hline
\multicolumn{10}{|c|}{}\\[-1em]
\multicolumn{10}{|c|}{$n_{\cool}$ versus $10^6 \times a^4 \chi$ $\qquad$ (SNF runs)}\\
%\multicolumn{10}{|c|}{}\\[-1em]
\hline
\multicolumn{10}{c}{}\\
\multicolumn{2}{c|}{$\beta=6.40$} & \multicolumn{2}{c|}{$\beta=6.50$} & \multicolumn{2}{c|}{$\beta=6.60$} & \multicolumn{2}{c|}{$\beta=6.70$} & \multicolumn{2}{c}{$\beta=6.80$} \\
\hline
10 &  4.94(24)  &  10 &  3.22(14) &  10 & 1.857(86) &  10 &  1.034(47) &  10 &  0.751(70) \\
20 &  5.10(24)  &  20 &  3.32(14) &  20 & 1.911(89) &  20 &  1.062(49) &  20 &  0.771(72) \\
30 &  5.16(25)  &  30 &  3.36(14) &  30 & 1.931(90) &  30 &  1.072(49) &  30 &  0.778(72) \\
40 &  5.20(25)  &  40 &  3.38(14) &  40 & 1.943(91) &  40 &  1.078(49) &  40 &  0.782(73) \\
50 &  5.22(25)  &  50 &  3.39(15) &  50 & 1.949(91) &  50 &  1.082(50) &  50 &  0.785(73) \\
60 &  5.24(25)  &  60 &  3.40(15) &  60 & 1.955(91) &  60 &  1.085(50) &  60 &  0.787(73) \\
\multicolumn{8}{c|}{} & 70 & 0.788(73) \\
\multicolumn{8}{c|}{} & 80 & 0.789(73) \\
%\hline
\end{tabular}
\caption{Summary of raw numerical data for the SNF runs. Each column represents $10^6 \times a^4 \chi$ as a function of $n_{\cool}$ for a given $\beta$.}
\label{tab:rawdata}
\end{minipage}
\hfill
\begin{minipage}{0.49\textwidth}
\footnotesize
\centering
\begin{tabular}{c|c|c|c|c}
Algorithm & $\beta$ & $n_{\cool}$ & $a/\sqrt{8t_0}$ & $(8t_0)^2 \chi$ \\
&&&&\\[-1em]
\hline
\multirow{5}{*}{Standard} & 5.95 & 6.8000 & 0.21568(22) & 0.03766(16) \\
& 6.00 & 8.1190 & 0.19739(14) & 0.03910(13) \\
& 6.07 & 10.203 & 0.17608(12) & 0.03989(14) \\
& 6.20 & 15.280 & 0.14388(15) & 0.04106(28) \\
& 6.40 & 26.811 & 0.10862(11) & 0.04095(28) \\
\hline
\multirow{5}{*}{SNF} & 6.40 & 26.811 & 0.10862(11) & 0.03698(177) \\
& 6.50 & 35.401 & 0.094527(85) & 0.04218(181) \\
& 6.60 & 46.132 & 0.082806(66) & 0.04142(193) \\
& 6.70 & 60.000 & 0.072608(77) & 0.03902(180) \\
& 6.80 & 77.753 & 0.063783(75) & 0.04765(444) \\
\end{tabular}
\caption{Summary of the data appearing in Fig.~\ref{fig:cont_lim_chi}. Standard simulation results come from Ref.~\cite{Bonanno:2025eeb}. Determinations of $t_0$ are interpolations of the data of Ref.~\cite{Giusti:2018cmp}.}
\label{tab:data_fig5}
\end{minipage}
\end{table}

\end{document}